\documentclass[aps,preprint]{revtex4}%
\usepackage{amsfonts}
\usepackage{amsmath}
\usepackage{amssymb}
\usepackage{graphicx}%
\providecommand{\U}[1]{\protect\rule{.1in}{.1in}}

\begin{document}
\title{A note on electrical and thermodynamic properties of Isolated Horizon}
\author{Gerui Chen}\email{chengerui@emails.bjut.edu.cn}
\affiliation{Institute of Theoretical Physics, Beijing University of
Technology, Beijing 100124, China}
\author{Xiaoning Wu}\email{wuxn@amss.ac.cn}
\affiliation{Institute of Mathematics, Academy of Mathematics and
System Science, Chinese Academy of Sciences, Beijing 100190,
China\\Hua Loo-Keng Key Laboratory of Mathematics, CAS, Beijing
100190, China\\State Key Laboratory of Theoretical Physics,
Institute of Theoretical Physics, Chinese Academy of Sciences,
Beijing 100190, China}
\author{Sijie Gao}\email{sijie@bnu.edu.cn}
\affiliation{Department of Physics, Beijing Normal University,
Beijing 100875, China}

\begin{abstract}
The electrical laws and Carnot cycle of Isolated Horizon (IH) are
investigated in this paper. We establish the Ohm's law and Joule's
law of an Isolated Horizon, and find that the conceptual picture of
black holes (Membrane Paradigm) can also apply to this kind of
quasi-local black holes. We also investigate the geometrical
properties near a non-rotating IH, and find that under the
first-order approximation of $r$, there exist a Killing vector and a
Hamiltonian conjugate to it, so this vector is a physical observer.
We calculate the energy as measured at infinity of a particle at
rest outside a non-rotating IH, and use this result to construct a
reversible Carnot cycle with the Isolated Horizon as a cold
reservoir, which confirms the thermodynamic nature of Isolated
Horizon.

\textbf{Keywords}: {Isolated Horizon, membrane paradigm, electrical
laws, black hole thermodynamics, reversible Carnot cycle}

\end{abstract}

\maketitle

\section{\textbf{Introduction}}

Since the first exact solution of Einstein equation was found out,
studying properties of black holes has become an important part of
gravitational physics. Black hole physics causes deep, unsuspected
connections among classical general relativity, quantum physics and
statistical mechanics. However, the classical definition of a black
hole \cite{gf,rmw}, which formalizes the notion of spacetime region
from which nothing can ever escape, is too global: it requires
knowledge of the entire future of the spacetime, and can not satisfy
the requirement for practical research \cite{aa7}, so alternate
notions of black hole \cite{dp,sa} were devised which possess a
quasi-local description formalism. In recent years, a new,
quasi-local framework was introduced by Ashtekar and his
collaborators to analyze different facets of black holes in a
unified way \cite{aa1,aa4,aa6,aa7}. Compared with the event horizon,
this framework doesn't need the knowledge of overall spacetime, and
only involves quasi-local conditions, so it accords with the
practical physical process. In this framework, black holes in
equilibrium are described by (Weakly) Isolated Horizons (WIH). This
paradigm leads to significant generalization of several results in
black hole physics and obtains considerable success.

In 1978, Damour found that instead of the conventional view of a
black hole as simply an empty region in spacetime, the black hole
horizon could be viewed as a physical membrane endowed with
specified mechanical and electromagnetic properties; for example,
the membrane of Kerr black hole behaves as a metallic shell whose
surface resistivity is $4\pi$ \cite{td}. Such idea was investigated
by other researchers later and was called ``membrane paradigm of
black hole" \cite{kt,mp}. This paradigm was reconsidered in the
framework of AdS/CFT correspondence more than twenty years later.
There are similar properties between the fluid/gravity duality in
the AdS/CFT framework and black-hole membrane \cite{gp} and this
kind of universality was interpreted as the Wilson renormalization
group flow in the AdS/CFT framework \cite{ni}. More developments in
this area can be found in Refs. \cite{ib}. This paper is concerned
with the electrical properties of Isolated Horizons. The electrical
laws of Kerr black hole \cite{td} are only related to the quantities
on the horizon, so it is natural to think that the laws are suitable
for Isolated Horizon.

Hawking radiation of Weakly Isolated Horizon was investigated
recently \cite{xnw1,xnw2}, which confirms one aspect of
thermodynamic nature of Isolated Horizon. Considering the importance
of Carnot cycle near a black hole to black hole thermodynamics
\cite{jdb,jdb2,uw}, we design a reversible Carnot cycle outside a
non-rotating Isolated Horizon to give a further confirmation of the
thermodynamic nature of Isolated Horizon.

The organization of this paper is as follows. In Section $2$, we
briefly review the definition and near-horizon geometry of (Weakly)
Isolated Horizon. In Section $3$, we establish the electrical laws
of an Isolated Horizon by following Damour's method. In Section $4$,
we investigate the properties near a non-rotating Isolated Horizon
under the first-order approximation of $r$, and construct a
reversible Carnot cycle near the horizon to confirm the
thermodynamical nature of Isolated Horizon. Finally, we make some
discussions and conclusions.

\section{Isolated Horizon and its near-horizon geometry}
In this section we briefly review the definition and geometric
properties of (Weakly) Isolated Horizon. According to the works by
Ashtekar and his collaborators \cite{aa1,aa4,aa6,aa7},
Weakly Isolated Horizon (WIH) is defined by\\
\textbf{Definition}  Let $ (M,g)$ be a spacetime. $H$ is a 3-dim
null hyper-surface in $M$ and $l^{a}$ is the tangent vector field of
the generator of $H$. $H$ is said to be a \textbf{Weakly
Isolated Horizon} (WIH), if\\
1)$H$ has the topology of $S^{2}\times R$;\\
2)The expansion of the null generator of $H$ is zero, i.e. $\Theta_{l}=0$ on $H$;\\
3)$T_{ab}\upsilon^{b}$ is future causal for any future causal vector
$\upsilon^{b}$ and Einstein equation holds in the neighborhood of
$H$;\\
4)$[\pounds_{l},D_{a}]l^{b}=0$ on $H$, where $D_{a}$ is the induced
covariant derivative on $H$.

A Weakly Isolated Horizon is said to constitute an \textbf{Isolated
Horizon} (IH) if $[\pounds_{l},D_{a}]=0$ on $H$ \cite{aa6}. From the
definition, Isolated Horizon is the special case of Weakly Isolated
Horizon, so IH has all the properties of WIH.

It is convenient to introduce Bondi-like coordinates
$(u,r,\theta,\varphi)$ in the neighborhood of the horizon $H$ in the
following way \cite{hf,bk}. First, denote the tangent vector of null
generator of $H$ as $l^{a}$ and another real null vector field as
$n^{a}$. The foliation of $H$ gives us the natural coordinates
$(\theta,\varphi)$. Lie dragging $(\theta,\phi)$ along each
generator of $H$ together with the parameter $u$ of $l^{a}$ forms
the coordinates $(u,\theta,\varphi)$ on $H$. Finally, choose the
affine parameter $r$ of $n^{a}$ as the fourth coordinate, then we
obtain the Bondi-like coordinates $(u,r,\theta,\phi)$ near the
horizon. With the Bondi-like coordinate system in hand, we construct
the null tetrad as
\begin{eqnarray}
l^{a}&=&\frac{\partial}{\partial u}+U\frac{\partial}{\partial
r}+X\frac{\partial}{\partial
\varsigma}+\overline{X}\frac{\partial}{\partial\overline{\varsigma}},\nonumber\\
n^{a}&=&-\frac{\partial}{\partial r},\nonumber\\
m^{a}&=&\omega \frac{\partial}{\partial
r}+\xi_{3}\frac{\partial}{\partial
\varsigma}+\xi_{4}\frac{\partial}{\partial \overline{\varsigma}},\nonumber\\
\overline{m}^{a}&=&\overline{\omega}\frac{\partial}{\partial
r}+\overline{\xi}_{3}\frac{\partial}{\partial
\overline{\varsigma}}+\overline{\xi}_{4}\frac{\partial}{\partial
\varsigma},\label{biaojia}
\end{eqnarray}
where $U\widehat{=}X\widehat{=}\omega\widehat{=}0 $ on $H$
(following the notation in Ref. \cite{aa7}, equalities restricted to
$H$ are denoted by ``$\widehat{=}$"), and $\varsigma
=e^{i\phi}cot\frac{\theta}{2}$. Note that $n^{a}$ and $l^{a}$ are
future directed.

We take the spacetime metric $g_{ab}$ to have a signature
$(-,+,+,+)$, so the metric can be expressed as
\begin{eqnarray}
g^{ab}=m^{a}\overline{m}^{b}+\overline{m}^{a}m^{b}-n^{a}l^{b}-l^{a}n^{b}.
\end{eqnarray}
The matrix form of the metric is
\begin{eqnarray}
  g^{\mu\nu}=\left(\begin{array}{cccc}
   0 & 1 & 0 & 0 \\
   1 & 2(U+|\omega|^{2}) & X+(\overline{\omega}\xi_{3}+\omega\overline{\xi}_{4}) &\overline{X}+(\overline{\omega}\xi_{4}+\omega\overline{\xi}_{3}) \\
   0 & X+(\overline{\omega}\xi_{3}+\omega\overline{\xi}_{4})& 2\xi_{3}\overline{\xi}_{4} &|\xi_{3}|^{2}+|\xi_{4}|^{2} \\
   0 & \overline{X}+(\overline{\omega}\xi_{4}+\omega\overline{\xi}_{3})
   &|\xi_{3}|^{2}+|\xi_{4}|^{2}&2\overline{\xi}_{3}\xi_{4}
 \end{array}\right),\label{rn6}
\end{eqnarray}
and, on the horizon, the metric reduces to
\begin{eqnarray}
 g^{\mu\nu}\widehat{=}\left(\begin{array}{cccc}
   0 & 1 & 0 & 0 \\
   1 & 0 & 0 & 0 \\
   0 & 0 &2\xi_{3}(0)\overline{\xi}_{4}(0) &|\xi_{3}(0)|^{2}+|\xi_{4}(0)|^{2} \\
   0 & 0
   &|\xi_{3}(0)|^{2}+|\xi_{4}(0)|^{2}&2\overline{\xi}_{3}(0)\xi_{4}(0)\label{1}
 \end{array}\right),
\end{eqnarray}
where $\xi_{3}(0)$ means the value of $\xi_{3}$ on the horizon. The
dual basis of the tetrad (\ref{biaojia}) can be calculated as
\begin{eqnarray}
\overline{m}_{a}&=&\frac{-\overline{\xi}_{3}}{|\xi_{4}|^{2}-|\xi_{3}|^{2}}d\varsigma+
\frac{\overline{\xi}_{4}}{|\xi_{4}|^{2}-|\xi_{3}|^{2}}d\overline{\varsigma}+
\frac{\overline{\xi}_{3}X-\overline{\xi}_{4}\overline{X}}{|\xi_{4}|^{2}-|\xi_{3}|^{2}}du,\nonumber\\
m_{a}&=&\frac{\xi_{4}}{|\xi_{4}|^{2}-|\xi_{3}|^{2}}d\varsigma+\frac{-\xi_{3}}{|\xi_{4}|^{2}-|\xi_{3}|^{2}}d\overline{\varsigma}+
\frac{\xi_{3}\overline{X}-\xi_{4}X}{|\xi_{4}|^{2}-|\xi_{3}|^{2}}du,\nonumber\\
-l_{a}&=&(U-\frac{\xi_{4}\overline{\omega}-
\overline{\xi}_{3}\omega}{|\xi_{4}|^{2}-|\xi_{3}|^{2}}X-\frac{\overline{\xi}_{4}\omega-\xi_{3}\overline{\omega}}{|\xi_{4}|^{2}-|\xi_{3}|^{2}}\overline{X})du-dr\nonumber\\
&&+\frac{\xi_{4}\overline{\omega}-\overline{\xi}_{3}\omega}{|\xi_{4}|^{2}-|\xi_{3}|^{2}}d\varsigma+
\frac{\overline{\xi}_{4}\omega-\xi_{3}\overline{\omega}}{|\xi_{4}|^{2}-|\xi_{3}|^{2}}d\overline{\varsigma},\nonumber\\
-n_{a}&=&du\label{badir1}.
\end{eqnarray}
The commutators--$[l^{a}, n^{a}]$ and $[m^{a}, n^{a}]$ tell us that
\begin{eqnarray}
\frac{\partial U}{\partial
r}&=&(\varepsilon+\overline{\varepsilon})+\overline{\pi}\overline{\omega}+\pi\omega
,\ \frac{\partial X}{\partial
r}=\overline{\pi}\overline{\xi}_{4}+\pi\xi_{3},\ \frac{\partial
\omega}{\partial
r}=\overline{\pi}+\overline{\lambda}\overline{\omega}+\mu\omega,\nonumber\\
\frac{\partial \xi_{3}}{\partial
r}&=&\overline{\lambda}\overline{\xi}_{4}+\mu\xi_{3},\
\frac{\partial\xi_{4}}{\partial
r}=\overline{\lambda}\overline{\xi}_{3}+\mu\xi_{4}\label{rn5}.
\end{eqnarray}

In the Newman-Penrose formalism, we can require the following gauge
conditions:
\begin{eqnarray}
\nu=\tau=\gamma=\alpha+\overline{\beta}-\pi=\mu-\overline{\mu}=0, \
\varepsilon-\overline{\varepsilon} \widehat{=} \kappa \widehat {=}
0,
\end{eqnarray}
which mean that the tetrad vectors (\ref{biaojia}) are parallelly
transported along $n^a$ in spacetime. The fourth requirement in the
definition of WIH implies that there exists a one form $\omega_{a}$
on $H$ such that $D_{a}l^{b}\widehat{=}\omega_{a}l^{b}$ and
$\pounds_{l}\omega^{a}\widehat{=}0$. In terms of the Newman-Penrose
formalism, $\omega_{a}$ can be expressed as
\begin{eqnarray}
\omega_{a}=-(\varepsilon+\overline{\varepsilon})n_{a}+(\alpha+\overline{\beta})\overline{m}_{a}+(\overline{\alpha}+\beta)m_{a}=-(\varepsilon+\overline{\varepsilon})n_{a}+\pi\overline{m}_{a}+\overline{\pi}m_{a},\label{bj6}
\end{eqnarray}
where $(\varepsilon+\overline{\varepsilon})$ is constant on $H$
\cite{aa7}. The definition of WIH also implies
\begin{eqnarray}
\rho\widehat{=}\sigma\widehat{=}0.
\end{eqnarray}

Based on Ref. \cite{aa7}, not any choice of time direction can give
a Hamiltonian evolution, and only some suitably chosen time
direction can lead to a well-defined horizon mass. In Ref.
\cite{aa7}, A. Ashtekar and B. Krishnan gave a canonical way to
choose the time direction $t^a$ for a WIH, and the restriction of
$t^a$ to $H$ should be a linear combination of the null normal $l^a$
and the axisymmetric vector $\psi^a$,
\begin{eqnarray}
t^a\ \widehat{=}\ B_tl^a-\Omega_{t}\psi^a\label{rn2},
\end{eqnarray}
where $B_t$ and $\Omega_{t}$ are constant on the horizon. Compared
with the Schwarzschild case, the parameter of $t^a$ takes the place
of the Killing time. With the canonical time direction $t^a$, Ref.
\cite{aa7} established the zeroth and the first law of WIH. By
definition, the surface gravity of $H$ is
$\kappa_{t}:=B_tl^a\omega_a=B_t(\varepsilon+\overline{\varepsilon})$,
which is constant on $H$, so the zeroth law of black hole mechanics
is valid for WIH. The first law is expressed as
\begin{eqnarray}
\delta M_{H}^{(t)}=\frac{\kappa_{t}}{8\pi}\delta
a_{H}+\Omega_{t}\delta J_{H}\label{jh},
\end{eqnarray}
where $M_{H}^{(t)} $ is the horizon mass, $ a_H $ is the area of the
cross section of WIH, $\Omega_{t}$ is the angular velocity of the
horizon, and $J_H=-\frac{1}{8\pi}\oint_S(\omega_a\psi^a)dS$ is the
angular momentum. The first law of WIH is the generalization of the
first law of stationary black hole. Since Hawking radiation of WIH
was investigated in Refs. \cite{xnw1,xnw2}, the laws of WIH
mechanics were upgraded to the laws of WIH thermodynamics.

\section{the electrical properties of Isolated Horizon}
In this section, we investigate the electrical laws of an Isolated
Horizon following Damour's method \cite{td}. Let us choose a set of
real Bondi-like coordinates $(u,r,\theta,\varphi)$. According to Eq.
(\ref{1}), and the fact that the topological structure of an
Isolated Horizons is $S^{2}\times R$, the metric on the horizon can
be also expressed as
\begin{eqnarray}
 g_{\mu \nu}\widehat{=}\left(\begin{array}{cccc}
   0 & 1 & 0 &0  \\
   1 & 0 & 0 &0  \\
   0 & 0 & f^{2}(\theta) &0  \\
   0 & 0 & 0 & f^{2}(\theta)\sin^{2}\theta
 \end{array}\right).\label{rn3}
\end{eqnarray}

The physical observer should be the canonical time $t^{a}$.
According to the relationship:
\begin{equation}
 (\frac{\partial}{\partial t})^{a}\widehat{=}B_{t}(\frac{\partial}{\partial
 u})^{a}-\Omega(\frac{\partial}{\partial \varphi})^{a},
\end{equation}
we introduce a new set of coordinates $(t
,\widetilde{r},\widetilde{\theta},\widetilde{\varphi})$. We have
substituted $\Omega$ for $\Omega_t$ for simplicity. The relationship
between Bondi-like coordinates $(u,r,\theta,\varphi)$ and the new
ones on the horizon is
\begin{equation}
t\widehat{=}\frac{1}{B_{t}}u,\ \widetilde{r}\widehat{=}r,\
\widetilde{\theta}\widehat{=}\theta,\
\widetilde{\varphi}\widehat{=}\frac{\Omega}{B_{t}}u+\varphi,
\end{equation}
so we have
\begin{eqnarray}
\frac{\partial}{\partial t}&=&\frac{\partial u}{\partial
t}\frac{\partial}{\partial u}+\frac{\partial r}{\partial
t}\frac{\partial}{\partial r}+\frac{\partial\theta}{\partial
t}\frac{\partial}{\partial \theta}+\frac{\partial \varphi}{\partial
t}\frac{\partial}{\partial
\varphi}\widehat{=}B_{t}\frac{\partial}{\partial
u}-\Omega\frac{\partial}{\partial \varphi},\nonumber\\
\frac{\partial}{\partial\widetilde{r}}&\widehat{=}&\frac{\partial}{\partial
r},\ \frac{\partial}{\partial
\widetilde{\theta}}\widehat{=}\frac{\partial}{\partial \theta},
\ \frac{\partial}{\partial\widetilde{\varphi}}\widehat{=}\frac{\partial}{\partial\varphi},\nonumber\\
dt&\widehat{=}&\frac{1}{B_{t}}du, \ d\widetilde{r}\widehat{=}dr, \
d\widetilde{\theta}\widehat{=}d\theta , \
d\widetilde{\varphi}\widehat{=}d\varphi+\frac{\Omega}{B_{t}}du.
\end{eqnarray}
The metric (\ref{rn3}) in the new coordinates can be expressed as
\begin{eqnarray}
 g_{\mu\nu}\widehat{=}\left(\begin{array}{cccc}
   0 & \frac{1}{B_{t}} & 0 &0  \\
   \frac{1}{B_{t}} & 0 & 0 &\frac{\Omega}{B_{t}} \\
   0 & 0 & f^{2}(\widetilde{\theta}) &0  \\
   0 & \frac{\Omega}{B_{t}} & 0 &
   f^{2}(\widetilde{\theta})\sin^{2}\widetilde{\theta}
 \end{array}\right)\label{ch1},
\end{eqnarray}
then the intrinsic geometry of cross section $t=$const and
$\widetilde{r}=0$ is
\begin{eqnarray}
d\widehat{s}^{2}=f^{2}(\widetilde{\theta})(d\widetilde{\theta}^{2}+\sin^{2}\widetilde{\theta}d\widetilde{\varphi}^{2}).
\end{eqnarray}
The basis of the cross section is
\begin{eqnarray}
e_{(\widetilde{\theta})}=\frac{1}{f}\frac{\partial}{\partial\widetilde{\theta}}
,\
e_{(\widetilde{\varphi})}=\frac{1}{f\sin\widetilde{\theta}}\frac{\partial}{\partial
\widetilde{\varphi}},
\end{eqnarray}
and the corresponding dual basis is
\begin{eqnarray}
\omega^{(\widetilde{\theta})}=fd\widetilde{\theta}, \
\omega^{(\widetilde{\varphi})}=f\sin\widetilde{\theta}d\widetilde{\varphi}.
\end{eqnarray}

Given an electromagnetic test field $F_{ab}$ regular on the horizon
$H$ we define the tangential electric field and the normal magnetic
induction by restricting the form
$F=\frac{1}{2}F_{\rho\sigma}(d\rho)_{a}\wedge(d\sigma)_{b}$ to the
horizon $\widetilde{r}=0$. Namely,
\begin{eqnarray}
F_{ab}&=&\frac{1}{2}F_{t
\sigma}(dt)_{a}\wedge(d\sigma)_{b}+\frac{1}{2}F_{\widetilde{r}\sigma}(d\widetilde{r})_{a}\wedge(d\sigma)_{b}+ \nonumber\\
&&\frac{1}{2}F_{\widetilde{\theta}\sigma}(d\widetilde{\theta})_{a}\wedge(d\sigma)_{b}+
\frac{1}{2}F_{\widetilde{\varphi}\sigma}(d\widetilde{\varphi})_{a}\wedge(d\sigma)_{b}\nonumber\\
&=&[F_{\widetilde{\theta}t}d\widetilde{\theta}+F_{\widetilde{\varphi}t}d\widetilde{\varphi}]\wedge
dt+F_{\widetilde{\theta}\widetilde{\varphi}}d\widetilde{\theta}\wedge
d\widetilde{\varphi}\nonumber\\
&=&(E_{(\widetilde{\theta})}\omega^{(\widetilde{\theta})}+E_{(\widetilde{\varphi})}\omega^{(\widetilde{\varphi})})\wedge
dt+B_{\perp}\omega^{(\widetilde{\theta})}\wedge\omega^{(\widetilde{\varphi})},
\end{eqnarray}
where
\begin{eqnarray}
E_{(\widetilde{\theta})}&=&F_{(\widetilde{\theta})t}=\frac{1}{f}F_{\widetilde{\theta}t}=
\frac{1}{f}(B_{t} F_{\theta u}-\Omega F_{\theta \varphi}),\nonumber\\
E_{(\widetilde{\varphi})}&=&F_{(\widetilde{\varphi})t}=
\frac{1}{f\sin\widetilde{\theta}}F_{\widetilde{\varphi}t}=\frac{1}{f\sin\theta}B_{t}F_{\varphi
u}=\frac{B_{t}}{f\sin\theta}F_{\varphi u},\nonumber\\
B_{\bot}&=&F_{(\widetilde{\theta})(\widetilde{\varphi})}=
\frac{1}{f^{2}\sin\widetilde{\theta}}F_{\widetilde{\theta}\widetilde{\varphi}}=\frac{1}{f^{2}\sin\theta}F_{\theta
\varphi}.\label{bj1}
\end{eqnarray}

By following Damour's method \cite{td}, we can endow charges and
currents on the horizon to keep the conservation of four-current
outside the black hole since we do not want to consider what happens
inside the black hole. There exists a four-current
$J(u,r,\theta,\varphi)$ which is defined and conserved all over
spacetime, $\nabla_{a}J^{a}=0$. We find a complementary current
$j^{a}$ with support on $r=0$ such that $J^{a}Y(r)+j^{a}$ is
conserved, where $Y$ is the Heaviside function. By replacing
$F^{ab}$ with $F^{ab}Y(r)$ in the equation
$J^{a}=\frac{1}{4\pi}\nabla_{b}F^{ab}$, we get the conserved current
$J^{a}Y(r)+j^{a}$ where $j^{a}=(4\pi)^{-1}F^{ar}\delta(r)$.

Next we calculate the Dirac distribution $\delta_{H}$ on the
horizon. For the spacetime, the volume element is
\begin{eqnarray}
\varepsilon_{abcd}&=&\varepsilon_{1234}\overline{m}_{a}\wedge
m_{b}\wedge l_{c}\wedge n_{d}=i\overline{m}_{a}\wedge m_{b}\wedge
l_{c}\wedge n_{d} \nonumber\\
&=&\frac{1}{|\xi_{4}|^{2}-|\xi_{3}|^{2}}\frac{\cot\frac{\theta}{2}}{\sin^{2}\frac{\theta}{2}}du\wedge
dr\wedge d\theta\wedge d\phi,
\end{eqnarray}
and the volume element of the cross section is
\begin{eqnarray}
\varepsilon_{ab}=i\overline{m}_{a}\wedge
m_{b}=\frac{1}{|\xi_{4}(0)|^{2}-|\xi_{3}(0)|^{2}}\frac{\cot\frac{\theta}{2}}{\sin^{2}\frac{\theta}{2}}d\theta\wedge
d\phi.
\end{eqnarray}
From the equation,
\begin{eqnarray} \int
f(u,r,\theta,\varphi)\delta_{H}\delta(u-u_{0})\varepsilon_{abcd}=
\int f(u_{0},0,\theta,\varphi)\varepsilon_{ab},
\end{eqnarray}
we easily find
\begin{eqnarray}
\delta_{H}=\delta(r).
\end{eqnarray}
Now we can write the complementary current $j^{a}$, with support on
the horizon, as
\begin{eqnarray}
j^{a}=K^{a}\delta_{H},
\end{eqnarray}
with
\begin{eqnarray}
K^{a}=\frac{1}{4\pi}F^{ar},
\end{eqnarray}
which is the surface four-current density and can be decomposed into
a surface charge density $\sigma$ and the geometrical components of
a surface current density $\overrightarrow{K}$,
\begin{eqnarray}
\sigma &=& K^{a}(dt)_{a}=\frac{1}{4 \pi B_{t}}F^{ur},\nonumber\\
K^{(\widetilde{\theta})}&=&\frac{1}{4\pi}F^{ar}\omega^{(\widetilde{\theta})}=\frac{1}{4\pi
f}F_{\theta u},\nonumber\\
K^{(\widetilde{\varphi})}&=&\frac{1}{4
\pi}F^{ar}\omega^{(\widetilde{\varphi})}=\frac{F_{\varphi u}}{4\pi
f\sin\theta}+V\sigma.\label{bj2}
\end{eqnarray}
It's not hard to find the following two relationships,
\begin{eqnarray}
\frac{1}{f}(B_{t}F_{\theta u}-\Omega F_{\theta
\varphi})+V_{(\widetilde{\varphi})}\frac{1}{f^{2}\sin\theta}F_{\theta
\varphi}=4\pi B_{t} \cdot \frac{1}{4\pi f}F_{\theta u},
\end{eqnarray}
\begin{eqnarray}
\frac{1}{f\sin\theta}B_{t}F_{\varphi u}=4\pi B_{t}
\cdot\frac{F_{\varphi u}}{4\pi f\sin\theta},
\end{eqnarray}
where $V_{(\widetilde{\varphi})}=f\sin\theta \Omega$ can be
interpreted as the velocity of the horizon. From Eqs. (\ref{bj1},
\ref{bj2}), we have
\begin{eqnarray}
E_{(\widetilde{\theta})}+V_{(\widetilde{\varphi})}B_{\bot}=4\pi
B_{t}K^{(\widetilde{\theta})},\nonumber\\
E_{(\widetilde{\varphi})}=4\pi
B_{t}[K^{(\widetilde{\varphi})}-\sigma V ].
\end{eqnarray}
In vector form,
\begin{eqnarray}
\overrightarrow{E}+\overrightarrow{V}\times\overrightarrow{B_{\bot}}=4\pi
B_{t}(\overrightarrow{K}-\sigma\overrightarrow{V}),\label{71}
\end{eqnarray}
which is the Ohm's law of  Isolated Horizon, and the surface
resistivity is $4\pi B_{t}$. For Kerr black holes,
$l^{a}\widehat{=}\partial_{t}+\Omega\partial_{\varphi}$, we have
$B_{t}=1$, so the surface resistivity is $4\pi$, which returns to
Damour's result \cite{td}.

It is useful to introduce the notions of the surface conduction
current $\overrightarrow{C}$(the total current $\overrightarrow{K}$
minus the convection current $\sigma\overrightarrow{V}$) and
the``dragged-along" electric field $\overrightarrow{E}^{\ast}$,
\begin{eqnarray}
\overrightarrow{C}=\overrightarrow{K}-\sigma\overline{V}, \
\overrightarrow{E}^{\ast}=\overrightarrow{E}+\overrightarrow{V}\times\overrightarrow{B_{\bot}}\label{rn4}.
\end{eqnarray}
So the Ohm's law (\ref{71}) can be rewritten as
\begin{eqnarray}
\overrightarrow{E}^{\ast}=4\pi B_{t}\overrightarrow{C}.
\end{eqnarray}

Next we  use the surface conduction current to express the Joule's
law. We can define the heat\,$dQ$ dissipated in the hole as
\cite{jdb}
\begin{eqnarray}
dQ=dM-\Omega dS_{z},
\end{eqnarray}
where $\Omega$ is the angular velocity and $dA, dM,$ and $dS_{z}$
are the increases in, respectively, area, mass, and angular momentum
of IH. The total energy flux into the hole is given by an integral
on the horizon \cite{swh},
\begin{eqnarray}
\dot{M}=\int T^{a}\!_{t}l_{a}dA=\int T_{at}l^{a}dA,
\end{eqnarray}
where $T_{ab}$ is the test energy-momentum tensor at the horizon.
The angular momentum flux is
\begin{eqnarray}
\dot{S}_{z}&=&
-\int T_{ab}l^{a}(\frac{\partial}{\partial
\widetilde{\varphi}})^{b}d A=-\int
T_{ab}l^{a}(\frac{\partial}{\partial \varphi})^{b}d A =-\int
T_{a\varphi}l^{a}dA.
\end{eqnarray}
So we get the heat production as
\begin{eqnarray}
\dot{Q}=\dot{M}-\Omega\dot{S_{z}}&=&\int(T_{at}+\Omega
T_{a\varphi} )l^{a}d A \nonumber\\
&=&B_{t}\int T_{ab}l^{a}l^{b}dA.
\end{eqnarray}
In the case of an electromagnetic field we have on the horizon,
\begin{eqnarray}
T_{ab}l^{a}l^{b}&=&(4\pi)^{-1}F_{ac}F_{b}\,^{c}l^{a}l^{b}=(4\pi)^{-1}F_{uc}F_{u}\,^{c}\nonumber\\
                &=&(4\pi)^{-1}(F_{ur}F_{u}\,^{r}+F_{u\theta}F_{u}\,^{\theta}+F_{u\varphi}F_{u}\,^{\varphi})\nonumber\\
                &=&(4\pi)^{-1}(g^{\theta\theta}F_{u\theta}F_{u\theta}+g^{\varphi\varphi}F_{u\varphi}F_{u\varphi})\nonumber\\
                &=&(4\pi)^{-1}(g^{\theta\theta}F_{\theta u}F_{\theta u}+g^{\varphi\varphi}F_{\varphi u}F_{\varphi
                u}),\label{bj4}
\end{eqnarray}
and from Eqs. (\ref{bj1}, \ref{bj2}), we have
\begin{eqnarray}
F_{\theta u}&=&\frac{1}{B_{t}}(fE_{(\widetilde{\theta})}+\Omega
F_{\theta \varphi})=\frac{f}{B_{t}}(E_{(\widetilde{\theta})}+VB_{\bot}),\nonumber\\
F_{\theta u}&=&4\pi fK^{(\widetilde{\theta})},\nonumber\\
F_{\varphi
u}&=&\frac{f\sin\theta}{B_{t}}E_{(\widetilde{\varphi})}=(K^{(\widetilde{\varphi})}-\sigma
V)4\pi f\sin\theta\label{bj3}.
\end{eqnarray}
By putting Eqs. (\ref{bj3}) into Eq. (\ref{bj4}), we obtain
\begin{eqnarray}
T_{ab}l^{a}l^{b}&=&\frac{1}{4\pi}\Big[\frac{1}{f^{2}}4\pi f
K^{(\widetilde{\theta})}\frac{f}{B_{t}}(E_{(\widetilde{\theta})}+VB_{\bot})\nonumber\\
                & &+\frac{1}{f^{2}\sin^{2}\theta}\frac{f\sin\theta}{B_{t}}E_{(\widetilde{\varphi})}4\pi
 f\sin\theta(K^{(\widetilde{\varphi})}-\sigma V )\Big]\nonumber\\
                &=&\frac{1}{B_{t}}\big[K^{(\widetilde{\theta})}(E_{(\widetilde{\theta})}+VB_{\bot})+E_{(\widetilde{\varphi})}(K^{(\widetilde{\varphi})}-\sigma V)\big]\nonumber\\
                &=&\frac{1}{B_{t}}(\overrightarrow{E}+\overrightarrow{V}\times\overrightarrow{B}_{\bot})\cdot(\overrightarrow{K}-\sigma\overrightarrow{V})\nonumber\\
                &=&\frac{1}{B_{t}}\overrightarrow{E}^{*}\cdot
                \overrightarrow{C},
\end{eqnarray}
where we have used Eqs. (\ref{rn4}) in the last equality.

Hence we get the Joule's law,
\begin{eqnarray}
\dot{Q}=B_{t}\int T_{ab}l^{a}l^{b}dA=\int 4\pi
B_{t}|\overrightarrow{C}|^{2}dA.
\end{eqnarray}
This is the Joule's law of Isolated Horizon, and the surface
resistivity is also $4\pi B_t$. From the two laws we see that IH can
be thought as a metal shell whose surface resistivity is $4\pi
B_{t}$. We do not use all the conditions of Isolated Horizon, so the
results are very general.

\section{Carnot cycle near a non-rotating Isolated Horizon}
In this section, we firstly investigate some properties near a
non-rotating IH, and find that under the first-order approximation
of $r$, $\frac{\partial}{\partial u}$ is a Killing vector and there
exists a Hamiltonian conjugate to it, so $\frac{\partial}{\partial
u}$ is a physical observer. Then we construct a reversible Carnot
cycle with the Isolated Horizon as a cold reservoir to confirm the
thermodynamic nature of IH.

For a non-rotating Isolated Horizon, the angular momentum is zero,
that is, $J_{H}=0$. According to the definition of angular momentum
of IH and Eq. (\ref{bj6}), we find that
\begin{eqnarray}
\pi\widehat{=}0,
\end{eqnarray}
so we have from the Eqs. (\ref{rn5}),
\begin{eqnarray}
\frac{\partial U}{\partial
r}\widehat{=}(\varepsilon+\overline{\varepsilon}), \ \frac{\partial
X}{\partial r}\widehat{=}0, \  \frac{\partial \omega}{\partial
r}\widehat{=}0.
\end{eqnarray}
The tetrad condition $\pounds_{l}m^{a}\widehat{=}0$ \cite{bk} means
that
\begin{equation} \frac{\partial \xi_{3}}{\partial
u}\widehat{=}0, \ \frac{\partial \xi_{4}}{\partial u}\widehat{=}0.
\end{equation}
For IH, $[\pounds_{l},D_{a}]\widehat{=}0$, so we have
$[\pounds_{l},D_{a}]n_b\widehat{=}0$. Considering
$\pounds_{l}n_a\widehat{=}0$, we obtain that
\begin{eqnarray}
\pounds_{l}(D_an_b)\equiv\pounds_{l}S_{ab}\widehat{=}0.
\end{eqnarray}
From Eqs. (B19, B20) in Ref. \cite{aa6},
\begin{eqnarray}
\widetilde{S}_{ab}:\widehat{=}\
\widetilde{q}^c_a\widetilde{q}^d_bD_cn_d\widehat{=}\mu(m_a\overline{m}_b+\overline{m}_am_b)+\lambda
m_am_b+\overline{\lambda}\overline{m}_a\overline{m}_b,\\
\pounds_{l}\widetilde{S}_{ab}\widehat{=}(D\mu)(m_a\overline{m}_b+\overline{m}_am_b)+(D\lambda)m_am_b+(D\overline{\lambda})\overline{m}_a\overline{m}_b,\label{rn7}
\end{eqnarray}
where $\widetilde{q}^a_b\ \widehat{=}\
m_b\overline{m}^a+\overline{m}_bm^a$, we find that
\begin{eqnarray}
\pounds_{l}\widetilde{S}_{ab}\widehat{=}\widetilde{q}^c_a\widetilde{q}^d_b\pounds_{l}S_{cd}\widehat{=}0,
\end{eqnarray}
so from the Eq. (\ref{rn7}), we have
\begin{eqnarray}
D\mu\widehat{=}0,\ D\lambda\widehat{=}0.
\end{eqnarray}
These two conditions above are given by Isolated Horizon. So we find
that $\frac{\partial U}{\partial r}, \ \frac{\partial X}{\partial
r}, \ \frac{\partial \omega}{\partial r}, \ \frac{\partial
\xi_{3}}{\partial r}, \ \frac{\partial \xi_{4}}{\partial r}$ are not
related to $u$ on the horizon, that is to say, the Taylor series of
$U, \ \omega, \ X, \ \xi_{3}, \ \xi_{4}$ to the first-order
approximation of $r$ are independent of $u$. From Eqs. (77, 78, 79)
in Ref. \cite{bk} and Eq. (\ref{rn6}) the metric and its inverse to
the first-order approximation of $r$ are not related to $u$, so
$\frac{\partial}{\partial u}$ can be thought as a Killing vector
approximatively.

Based on Ref. \cite{rmwaz}, the necessary and sufficient condition
for the existence of a Hamiltonian conjugate to $\eta^a$ on slice
$\Sigma$ is
\begin{eqnarray}
\int_{\partial \Sigma}\eta \cdot \omega(g_{ab}, \delta_{1}g_{ab},
\delta_{2}g_{ab})=0,
\end{eqnarray}
where $\omega$ is the presymplectic current 3-form. In general
relativity, the presymplectic current 3-form is
\begin{eqnarray}
\omega_{abc}=\frac{1}{16\pi}\epsilon_{dabc}w^d,
\end{eqnarray}
where
\begin{eqnarray}
w^a=P^{abcdef}[\gamma_{2bc}\nabla_{d}\gamma_{1ef}-\gamma_{1bc}\nabla_{d}\gamma_{2ef}],
\end{eqnarray}
with
\begin{eqnarray}
P^{abcdef}=g^{ae}g^{fb}g^{cd}-\frac{1}{2}g^{ad}g^{be}g^{fc}-\frac{1}{2}g^{ab}g^{cd}g^{ef}-\frac{1}{2}g^{bc}g^{ae}g^{fd}+\frac{1}{2}g^{bc}g^{ad}g^{ef}.
\end{eqnarray}
In our case it is straightforward to find that
\begin{eqnarray}
\int\frac{\partial}{\partial u} \cdot \omega(g_{ab},
\delta_{1}g_{ab}, \delta_{2}g_{ab})&=&\int(\frac{\partial}{\partial
u})^l\omega_{ljk}=\int(\frac{\partial}{\partial
u})^l\epsilon_{aljk}\frac{w^a}{16\pi}\nonumber\\
&=&\int(\frac{\partial}{\partial u})^l i\overline{m}_{a}\wedge
m_{l}\wedge l_{j}\wedge n_{k}\frac{w^a}{16\pi}\thicksim O(r^2).
\end{eqnarray}
In the last equality, we use $(\frac{\partial}{\partial
u})^lm_l\thicksim O(r^2)$. So there exists a Hamiltonian conjugate
to $\frac{\partial}{\partial u}$ near the horizon, and
$\frac{\partial}{\partial u}$ can be thought as a physical observer.

Next we will compute the energy as measured at infinity of a
particle near the horizon. The Taylor series of the coefficients to
the first-order approximation of $r$ are
\begin{eqnarray}
U&=&(\varepsilon+\overline{\varepsilon})r+O(r^{2}),\nonumber\\
X&=&O(r^{2}),\nonumber\\
\omega&=&O(r^{2}),\nonumber\\
\xi_{3}&=&\xi_{3}(0)+\xi_{3}^{(1)}(0)r+O(r^{2}),\nonumber\\
\xi_{4}&=&\xi_{4}(0)+\xi_{4}^{(1)}(0)r+O(r^{2}),\label{hr1}
\end{eqnarray}where $\xi_{3}^{(1)}$ is the first-order
derivative of $r$. Putting expressions (\ref{hr1}) into the null
tetrad (\ref{biaojia}), we get
\begin{eqnarray}
l^{a}&=&\frac{\partial}{\partial
u}+\big[(\varepsilon+\overline{\varepsilon})r+O(r^{2})\big]\frac{\partial}{\partial
r},\nonumber\\
n^{a}&=&-\frac{\partial}{\partial r},\nonumber\\
m^{a}&=&O(r^{2})\frac{\partial}{\partial
r}+\big[\,\xi_{3}(0)+\xi_{3}^{(1)}(0)r+O(r^{2})\big]\frac{\partial}{\partial
\varsigma}\nonumber\\
&&+\big[\,\xi_{4}(0)+\xi_{4}^{(1)}(0)r+O(r^{2})]\frac{\partial}{\partial
\overline{\varsigma}},\nonumber\\
\overline{m}^{a}&=&O(r^{2})\frac{\partial}{\partial
r}+\big[\,\overline{\xi_{3}}(0)+\overline{\xi_{3}}^{(1)}(0)r+O(r^{2})\big]\frac{\partial}{\overline{\varsigma}}\nonumber\\
&&+\big[\,\overline{\xi_{4}}(0)+\overline{\xi_{4}}^{(1)}(0)r+O(r^{2})\big]\frac{\partial}{\partial
\varsigma}.\label{jiehe1}
\end{eqnarray}

Introduce new coordinates in the neighborhood of IH as \cite{xnw2}
\begin{eqnarray}
t:=\frac{1}{B_t}u-r_{*}, R:=r ,
\end{eqnarray}
where $dr_{*}=\frac{dr}{f(r)}$ and
$f(r)=2B_t(\varepsilon+\overline{\varepsilon})r$. It is easy to find
that
\begin{eqnarray}
\frac{\partial}{\partial u}&=&\frac{1}{B_t}\frac{\partial}{\partial
t} ,\ \frac{\partial}{\partial r}=-\frac{1
}{f}\frac{\partial}{\partial t}+\frac{\partial}{\partial R},\label{31}\\
du&=&B_tdt+B_t\frac{dR}{f(R)},\  dr=dR.\label{hr2}
\end{eqnarray}
We use Eqs. (\ref{31}) to express the null tetrad (\ref{jiehe1}) in
the new coordinates as
\begin{eqnarray}
l^{a}&=&\frac{1}{2B_t}\frac{\partial}{\partial
t}+[(\varepsilon+\overline{\varepsilon})R+O(R^{2})]\frac{\partial}{\partial
R},\nonumber\\
n^{a}&=&\frac{1}{f}\frac{\partial}{\partial
t}-\frac{\partial}{\partial R},\nonumber\\
m^{a}&=&O(R^{2})\big[-\frac{1}{f}\frac{\partial}{\partial
t}+\frac{\partial}{\partial
R}\big]+\big[\xi_{3}(0)+\xi_{3}^{(1)}(0)R+O(R^{2})\big]\frac{\partial}{\partial
\varsigma}\nonumber\\
&&+\big[\xi_{4}(0)+\xi_{4}^{(1)}(0)R+O(R^{2})\big]\frac{\partial}{\partial
\overline{\varsigma}},\nonumber\\
\overline{m}^{a}&=&O(R^{2})\big[-\frac{1}{f}\frac{\partial}{\partial
t}+\frac{\partial}{\partial
R}\big]+\big[\,\overline{\xi_{3}}(0)+\overline{\xi_{3}}^{(1)}(0)R+O(R^{2})\big]\frac{\partial}{\partial
\overline{\varsigma}}\nonumber\\
&&+\big[\,
\overline{\xi_{4}}(0)+\overline{\xi_{4}}^{(1)}(0)R+O(R^{2})\big]\frac{\partial}{\partial
\varsigma}.
\end{eqnarray}
From the expressions above we get easily
\begin{eqnarray}
g^{tt}\dot{=}-\frac{1}{B_tf(R)}.
\end{eqnarray}
Putting  Eqs. (\ref{hr2}) into Eqs. (\ref{badir1}) and ignoring the
higher order terms, we obtain
\begin{eqnarray}
g_{RR}&=&-\frac{2B_t^2}{f^{2}(R)}(U-\frac{\xi_{4}\overline{\omega}-
\overline{\xi}_{3}\omega}{|\xi_{4}|^{2}-|\xi_{3}|^{2}}X-\frac{\overline{\xi}_{4}\omega-
\xi_{3}\overline{\omega}}{|\xi_{4}|^{2}-|\xi_{3}|^{2}}\overline{X})\nonumber\\
&&+\frac{2B_t^2}{f^{2}(R)}\frac{(\xi_{3}\overline{X}-\xi_{4}X)(\overline{\xi_{3}}X-\overline{\xi_{4}}\overline{X})}{(|\xi_{4}|^{2}-|\xi_{3}|^{2})^{2}}+\frac{2B_t}{f(R)}\nonumber\\
&\dot{=}&\frac{B_t}{f(R)},\nonumber\\
g_{tR}&\dot{=}&0,\nonumber\\
g_{tt}&\dot{=}&-B_tf(R).
\end{eqnarray}
We will use these results later.

The Hamilton-Jacobi equation in curved spacetime is
\begin{eqnarray}
g^{\mu\nu}\partial_{\mu} S\partial_{\nu} S+m^{2}=0,\label{2}
\end{eqnarray}
where $m$ is the rest mass, and $S$ is the Hamilton principal
function. Because $\frac{\partial}{\partial
t}=B_t\frac{\partial}{\partial u}$ is a Killing vector, we can
separate $S$ as
\begin{eqnarray}
S=-V(t)+W(R,x^{A}).
\end{eqnarray}
The components of  generalized momentum $P_{\mu}=\frac{\partial
S}{\partial x^{\mu}}$ are
\begin{eqnarray}
P_{t}&=&\frac{\partial S}{\partial t}=-\frac{\partial V(t)}{\partial
t}=-\dot{V}(t),\nonumber\\
P_{R}&=&\frac{\partial S}{\partial R}=\frac{\partial
W(R,x^A)}{\partial
R},\nonumber\\
P_{A}&=&\frac{\partial W(R,x^{A})}{\partial x^{A}},
\end{eqnarray}
where $\dot{V}(t)$ is the energy of the particle.
 We consider a
particle at rest outside a non-rotating Isolated Horizon, so for
this particle, $R$ and $x^A$ are constant, which means
\begin{eqnarray}
P_{R}=0, \ P_{A}=0\label{4}.
\end{eqnarray}
Then Eq. (\ref{2}) becomes
\begin{eqnarray}
g^{tt}(\dot{V})^{2}+m^{2}=0,
\end{eqnarray}
and it is easy to find that
\begin{eqnarray}
\dot{V}=\sqrt{B_tfm^{2}}.\label{10}
\end{eqnarray}
This is the energy as measured at infinity of a particle at rest
outside the horizon. When a particle is located at $R=\delta$ which
is the radial coordinate distance away from the horizon, we have
\begin{eqnarray}
f=2B_t(\varepsilon+\overline{\varepsilon})\delta.
\end{eqnarray}
Putting this expression into Eq. (\ref{10}), we obtain
\begin{eqnarray}
\dot{V}=\sqrt{2(\varepsilon+\overline{\varepsilon})}\delta^{\frac{1}{2}}mB_t.\label{zhs1}
\end{eqnarray}

Now let us change the coordinate distance $\delta$ to the proper
distance $l$. The general express is
\begin{eqnarray}
d\l^{2}=\gamma_{ij}dx^{i}dx^{j}=(g_{ij}-\frac{g_{0i}g_{0j}}{g_{00}})dx^{i}dx^{j}.
\end{eqnarray}
We consider the radial distance, so
\begin{eqnarray}
d\l^{2}=g_{RR}dR^{2}=\frac{B_t}{f(R)}dR^{2},\label{14}
\end{eqnarray}
that is,
\begin{eqnarray}
d\l=\left[\frac{B_t}{f(R)}\right]^{\frac{1}{2}}dR.
\end{eqnarray}
So the proper distance $l$ corresponding to  coordinate distance
$\delta$ is
\begin{eqnarray}
l&=&\int_{0}^{\delta}\left[\frac{B_t}{f(R)}\right]^{\frac{1}{2}}dR=\int_{0}^{\delta}\left[\frac{1}{2(\varepsilon+\overline{\varepsilon})R}\right]^{\frac{1}{2}}dR\nonumber\\
&=&\frac{1}{[2(\varepsilon+\overline{\varepsilon})]^{\frac{1}{2}}}\int_{0}^{\delta}\left[\frac{1}{R}\right]^{\frac{1}{2}}dR=\frac{1}{[2(\varepsilon+\overline{\varepsilon})]^{\frac{1}{2}}}\int_{0}^{\delta}R^{-\frac{1}{2}}dR\nonumber\\
&=&\frac{\sqrt{2}}{(\varepsilon+\overline{\varepsilon})^{\frac{1}{2}}}\delta^{\frac{1}{2}}.\label{12}
\end{eqnarray}
Putting the result above into Eq. (\ref{zhs1}), we have
\begin{eqnarray}
\dot{V}=m lB_t(\varepsilon+\overline{\varepsilon}).
\end{eqnarray}
The surface gravity of IH is
$\kappa=B_t(\varepsilon+\overline{\varepsilon})$, so the energy can
be rewritten as
\begin{eqnarray}
\dot{V}=m\cdot l \cdot \kappa.
\end{eqnarray}
This is the energy as measured at infinity of a particle at rest
outside the horizon with the radial proper distance $l$ away from
the horizon.

The red-shift factor is defined by
\begin{eqnarray}
\chi=\sqrt{-g_{00}},
\end{eqnarray}
and we have
\begin{eqnarray}
\frac{d\chi}{dl}=\frac{d\chi}{dR}\frac{dR}{dl}=\frac{d\sqrt{-g_{00}}}{dR}\frac{1}{\sqrt{g_{11}}}=B_t(\varepsilon+\overline{\varepsilon})=\kappa,
\end{eqnarray}
so the energy as measured at infinity of the particle can be
expressed as
\begin{eqnarray}
\dot{V}=m\chi.
\end{eqnarray}
This is the important result we will use in the following
discussion.

Hawking radiation of Isolated Horizon was confirmed recently
\cite{xnw1,xnw2}, however, can it really be regarded as a
thermodynamical system? We will analyze its thermodynamical property
in another way. Following Ref. \cite{dxh}, we design a reversible
Carnot cycle with a non-rotating Isolated Horizon being the cold
reservoir. The thermodynamic nature of Isolated Horizon is confirmed
further in the framework of Carnot cycle.

We divide the total process into the following four steps (see Fig.
1):

\textbf{Step (a)} A hot reservoir near an Isolated Horizon is filled
with thermal radiation of temperature $T_{1}$. Like Ref. \cite{dxh}
we require the chemical potential of the thermal radiation vanish.
The Isolated Horizon is used as the cold reservoir, so we need
$T_{1}>T_{H}$ ($T_H$ is the temperature of IH). The initial state of
the working substance is an empty box and the mass of the empty box
is negligible. For designing a reversible process, we replace one
side of the box with a piston in our model. When we pull the piston
slowly down to the other side of the box such that the box finally
is full of thermal radiation of temperature $T_{1}$, there is some
work done by the thermal radiation (see Fig. 1). After the process
above, we have a box of substance with energy $E_{1}$. Since the
process step (a) is quasi-static and isothermal, we can use the
first law of thermodynamics in the following form,
\begin{eqnarray}
Q_{1}^{\prime}=T_{1}S_{1}=E_{1}+W_{1}^{\prime},
\end{eqnarray}
where $Q_{1}^{\prime}$ is the heat flowing into the box, $S_{1}$ is
the entropy of the matter in the box, and $W_{1}^{\prime}$ is the
work done by the thermal radiation. Please note that all the
quantities are locally measured, so the corresponding quantities
measured at infinity are
\begin{eqnarray}
Q_{1}=\chi_{1}T_{1}S_{1}=\chi_{1}E_{1}+\chi_{1}W_{1}^{\prime}=\chi_{1}E_{1}+W_{1},\label{26}
\end{eqnarray}
where $\chi_{1}$ is the red-shift factor at the hot reservoir.\\
\begin{figure}[htmb]
\centering \scalebox{1.1} {\includegraphics{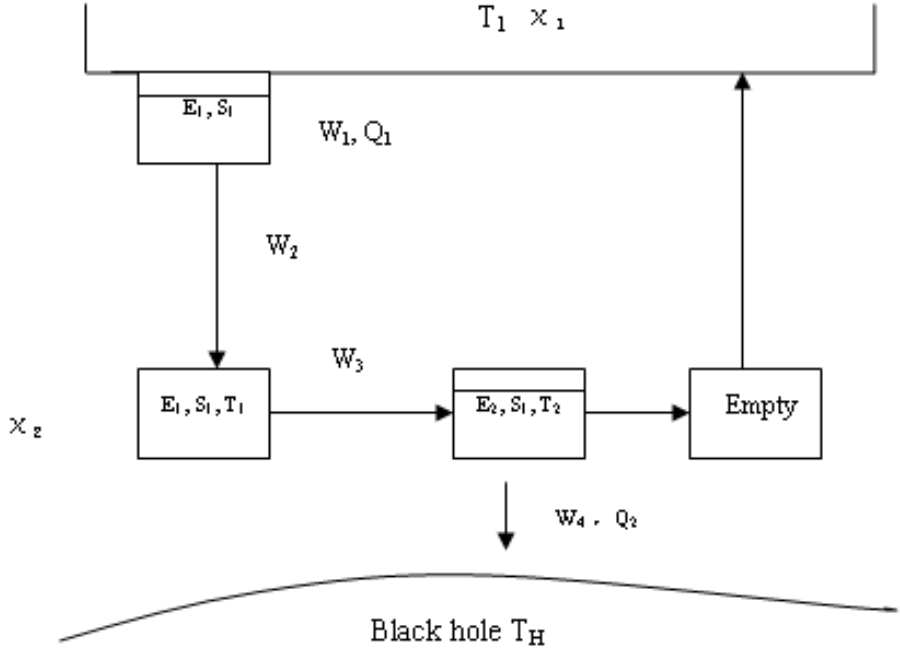}} \caption{a
reversible Carnot Cycle outside a non-rotating Isolated Horizon}
\label{map}
\end{figure}

\textbf{Step (b)} adiabatic process. This process actually consists
of two sub-steps.

\textbf{Sub-step (b1)}The box is lowered slowly down to some point
near the black hole with the red-shift factor $\chi_{2}$. In this
step, the substance inside does not change, so all locally measured
quantities, such as energy, entropy, temperature, etc, remain
unchanged. However, the energy measured by the agent holding the
string at infinity becomes $\chi_{2}E_{1}$, which means the work
done in the lowering process is
\begin{eqnarray}
W_{2}=E_{1}(\chi_{1}-\chi_{2}).
\end{eqnarray}
Please note that this is not the thermodynamic work done by the
substance in the box since the volume of the substance is unchanged,
and the work is done by gravity in a pure mechanical process. An
analog of the adiabatic process in an ordinary Carnot cycle is given
in sub-step (b2) below.

\textbf{Sub-step (b2)} This sub-step plays an essential role in our
model. In a typical Carnot cycle, the working substance should
experience an adiabatic expansion to cool down to temperature
$T_{2}$, the temperature of the cold reservoir. However, the
adiabatic lowering process--sub-step (b1) does not change the
temperature of the substance, so it is necessary to add another
adiabatic process to cool down the substance from $T_{1}$ to
$T_{2}$. But the temperature at any point outside the black hole is
not well-defined since the box is not in direct contact with the
Isolated Horizon. We can use a plausible criterion, i.e. the
conservation of the total entropy, to specify $T_{2}$ as follows.
Suppose that the box cools down to $T_{2}$ via an adiabatic
expansion. The entropy of box remains
$S_{1}=\frac{Q_{1}}{\chi_{1}T_{1}}$ (see Eq. (\ref{26})). Then the
box gives up all its energy to the black hole which is shown in step
(c) below. The energy that the black hole finally absorbs is
$Q_{2}=\chi_{2}T_{2}S_{1}$ (see Eq. (\ref{25})). Hence the change in
the total entropy is
\begin{eqnarray}
\triangle S=-S_{1}+\frac{\chi_{2} T_{2}S_{1}}{T_{H}}\label{by1}.
\end{eqnarray}
Because for a reversible adiabatic process, the total entropy should
not change, that is, $\Delta S=0$, we find immediately from Eq.
(\ref{by1}),
\begin{eqnarray}
T_{2}=\frac{T_{H}}{\chi_{2}}.
\end{eqnarray}

Different from sub-step (b1), sub-step (b2) changes the parameter
set of the box from $(E_{1}, T_{1})$ to $(E_{2}, T_{2})$. The
entropy $S_{1}$ keeps unchanged and the adiabatic expansion takes
place in the same location, so the work measured from infinity is
given by
\begin{eqnarray}
W_{3}=\chi_{2}(E_{1}-E_{2}).
\end{eqnarray}

\textbf{Step (c)} Release the substances from the box into the black
hole. We use the piston in the opposite way to push out the thermal
radiation in a quasi-static and isothermal manner. We have
\begin{eqnarray}
Q_{2}^{\prime}=T_{2}S_{1}=E_{2}-W_{4}^{\prime},
\end{eqnarray}
where prime is used to label locally measured quantity. The
corresponding quantities measured at infinity are
\begin{eqnarray}
Q_{2}&=&\chi_{2}Q_{2}^{\prime}=\chi_{2}T_{2}S_{1},\label{25}\\
W_{4}&=&\chi_{2}W_{4}^{\prime}=\chi_{2}(E_{2}-Q_{2}^{\prime}).
\end{eqnarray}
Note that $Q^{\prime}_{2}$ is the energy released from the box seen
by a local observer while the energy absorbed by the black hole is
$Q_{2}$, instead of $Q^{\prime}_{2}$.

\textbf{Step (d)} Lift the empty box back and return to its initial
state. Nothing happens in this process because the mass of the empty
box is negligible.

Now let us compute the thermal efficiency for the complete
reversible Carnot cycle. It is easy to check that the total work
$W\equiv W_{1}+W_{2}+W_{3}+W_{4}$ satisfies the familiar relation in
a typical Carnot cycle,
\begin{eqnarray}
W=Q_{1}-Q_{2}.
\end{eqnarray}
Hence the efficiency is
\begin{eqnarray}
\eta&=&\frac{W}{Q_{1}}=1-\frac{Q_{2}}{Q_{1}}\nonumber\\
 &=&1-\frac{\chi_{2}}{\chi_{1}}\frac{T_{2}}{T_{1}}\nonumber\\
 &=&1-\frac{T_{H}}{\chi_{1}T_{1}}.\label{27}
\end{eqnarray}
This is the desired efficiency for a reversible Carnot engine
operating between two heat sources with temperatures $T_{1}$ and
$T_{H}$ respectively. Therefore the thermodynamic nature of Isolated
Horizon is confirmed in the framework of Carnot cycle and the
efficiency of this cycle confirms that Isolated Horizon behaves as a
thermodynamic object with Hawking temperature $T_H$.

\section{conclusions and discussions}
In this paper, we investigate the electrical and thermodynamical
properties of Isolated Horizon. By following Damour's method, we
establish the Ohm's law and Joule's law of an Isolated Horizon, so
we generalize Damour's results. From the calculation, we find that
the results are very general, since we do not use all the conditions
of Isolated Horizon. We investigate the geometry in the vicinity of
a non-rotating Isolated Horizon, and find that under the first-order
approximation of $r$, $\frac{\partial}{\partial u}$ is a Killing
vector and there exists a Hamiltonian conjugate to it, so
$\frac{\partial}{\partial u}$ is a physical observer. We calculate
the energy as measured at infinity of a particle at rest outside the
horizon and construct a reversible Carnot Cycle with an Isolated
Horizon as a cold reservoir, which gives a further confirmation of
the thermodynamic nature of Isolated Horizon.

\begin{acknowledgements}
This research was supported by NSFC Grants No. 11175245, 11075206, 11235003, 11375026 and
NCET-12-0054.

\end{acknowledgements}

\end{document}